\documentclass[twoside]{article}
\usepackage{fleqn,espcrc2}

\newcommand{\AmS}{{\protect\the\textfont2
  A\kern-.1667em\lower.5ex\hbox{M}\kern-.125emS}}

\title{Division Algebras, Extended Supersymmetries and Applications}

\author{F. Toppan\address[MCSD]{CCP - CBPF,
        Rua Dr. Xavier Sigaud 150, \\
        cep 22290-180, Urca, Rio de Janeiro (RJ), Brazil}%
        }

\begin{document}

\begin{abstract}
I present here some new results which make explicit the role of
the division algebras ${\bf R}$, ${\bf C}$, ${\bf H}$, ${\bf O}$
in the construction and classification of, respectively,
$N=1,2,4,8$ global supersymmetric quantum mechanical and classical
dynamical systems. In particular an $N=8$  Malcev superaffine
algebra is introduced and its relation to the non-associative
$N=8$ SCA is discussed. A list of present and possible future
applications is given.
 \vspace{1pc}
\end{abstract}

\maketitle

\section{INTRODUCTION}

The investigation of extended supersymmetries in quantum
mechanical systems is an essential tool for any realistic
application of supersymmetry. Many examples can be given. Here I
limit myself just to mention that any supersymmetric field theory
in the ordinary Minkowski space, when dimensionally reduced to a
one-temporal dimension \cite{ClHa}, acquires $4$ times the number
of the original supersymmetries. For instance $N=1,2,4$
super-YangMills theories are reduced to $N=4,8,16$ super quantum
mechanical systems respectively.\par Due to the lack of an
explicit superfield formalism for large $N$ ($N>4$), such systems
have received little attention in the literature, apart some
specific cases, see e.g. \cite{CoPa} and \cite{GaRa}.\par On the
other hand, since at least the work of Kugo and Townsend
\cite{KuTo}, it is known that the division algebras are related
with the $N=1,2,4,8$ extended supersymmetries. However, still
nowadays, the connection between division algebras and extended
supersymmetries has not been completely worked out. In several
cases it has been made explicit; in many other cases however it is
only implicit or just suggested. This is especially true for the
division algebra of octonions (or octaves) since, being
non-associative, is technically more difficult to handle.
\par The aim of this paper is to partly fill this gap,
presenting some new constructions which illustrate and make
explicit such a connection in a few new cases. The results here
presented are extracted from three recent articles \cite{PaTo},
\cite{CRT} and \cite{CRT2}. Their content being more general, only
the part concerning the role of division algebras is here
discussed. Proofs and details can be found there.\par More
specifically, the points here discussed are the relation between
octonions and a given real Clifford algebra associated to the
representation of the $N=8$ supersymmetric quantum mechanics
algebra, as well as the explicit construction of the global $N=8$
supersymmetry in terms of charges derived from a super-current
algebra which turns out to be a Malcev algebra (i.e. a special
type of non-jacobian algebra). An algebraic interpretation (via
Sugawara costruction) for the so-called non-associative $N=8$
superconformal algebra introduced in \cite{ESTvPS} is obtained as
an extra bonus.\par The above construction finds an immediate
application in the extended supersymmetrization of integrable
equations in $1+1$ dimensions (such as KdV) \cite{CRT2}. A list of
possible other applications is given in the Conclusions.

\section{DIVISION ALGEBRAS AND EXTENDED SUPERSYMMETRIES}

Let me recall some basic properties of the division algebras ${\bf
R}, {\bf C}, {\bf H}, {\bf O}$ and of their relation with extended
supersymmetries.\par Division algebras over the real field ${\bf
R}$ can be obtained by iterating the ``doubling" procedure (see
\cite{Pos}) which allows to construct complex numbers out of real
numbers.\par After each doubling some properties are lost.
Quaternions are non-commutative, while octonions are
non-associative. They satisfy however the weaker condition of
alternativity (i.e. the associator $\relax [a,b,c]\equiv
(ab)c-a(bc)$ vanishes for $a=b$).\par By virtue of the generalized
Frobenius theorem, ${\bf R}, {\bf C}, {\bf H}, {\bf O}$ are the
only finite, alternative algebras over the real field up to
isomorphism. In this respect they are a very fundamental and
peculiar mathematical structure.\par They can be used to construct
an explicit realization of the one-dimensional supersymmetry
algebra
\begin{eqnarray}
\{ Q_i, Q_j\} &=& \delta_{ij} H, \label{susy}
\end{eqnarray}
where the total number of supersymmetry charges $Q_i$ is $1,2,4$
or $8$ respectively.\par The relation can be made precise as
follows. Let $\tau_a$ denote the imaginary octonions (for
$a=1,2,...,7$, while imaginary quaternions are restricted to the
values $a=1,2,3$, and the imaginary unit is assumed to be given
for $a=1$).\par Explicitly, their multiplication rule can be
presented through
\begin{eqnarray}
\tau_a\cdot\tau_b &=& -\delta_{ab}{\bf 1} + C_{abc}\tau_c,
\end{eqnarray}
where the structure constants $C_{abc}$ can be assumed to be
totally antisymmetric and non-vanishing only for the special
combinations \begin{eqnarray} abc\equiv 123, 147, 165, 246, 257,
354, 367 \label{combination} \end{eqnarray} ($C_{abc}$ is
normalized to $1$ for even permutations of the above
orderings).\par If we introduce an $N=1$ one-dimensional
superspace with bosonic coordinate $x$ and Grassmann variable
$\theta$, we can easily construct an ($N=2,4,8$) extended version
of the supersymmetry algebra in terms of the generators
\begin{eqnarray}
Q_{0} \equiv Q &=& \frac{\partial}{\partial\theta}
+\theta\frac{\partial}{\partial x}, \label{transf1}
\end{eqnarray}
while
\begin{eqnarray}
Q_a\equiv \tau_a D, \label{transf2}
\end{eqnarray}
with
\begin{eqnarray}
D &=&\frac{\partial}{\partial\theta}
-\theta\frac{\partial}{\partial x}, \label{derivative}
\end{eqnarray}
(here $i\equiv 0, a$). The introduction of imaginary octonions has
been used to ``correct" the sign appearing in the anticommutation
of the fermionic derivative.

\section{$N=8$ REPRESENTATION MULTIPLETS AND OCTONIONS}

Being non-associative, octonions cannot of course be represented
through matrices with the standard product. On the other hand it
is well-known that octonions and pseudo-octonions are related to
certain real representations of the Clifford algebras $C(p,q)$
(i.e. for a $(p,q)$ signature), such as $C(0,7)$, $C(0,8)$ and
$C(8,0)$. An explicit construction is given in \cite{Oku}.
\par
The $16$-dimensional matrix representations $C(8,0)$ and $C(0,8)$
are of particular interest in the light of the results of
\cite{PaTo}. In that paper some properties of the finite
irreducible multiplets of representation of the one-dimensional
$N$-extended supersymmetry have been proven. The results can be
summarized as follows. At first it is shown that any such
multiplet contains an equal number (say $d$) of bosons and
fermions distributed in different spin states. Any given multiplet
under consideration can be unambiguously ``shortened" to a
standard (fundamental) multiplet such that all $d$ bosons and all
$d$ fermions are grouped in just two spin states (one for the
bosons, the other one for the fermions). Since all irreducible
representations are specified by the given irreducible fundamental
multiplet, which can be regarded as both the generator and the
representative of its class, it is of tantamount importance to
classify all classes of irreducible fundamental multiplets. The
answer can be given rather easily. It turns out that they are in
one-to-one correspondence with the real representations of
Clifford algebras which are of Weyl form, i.e. such that all
matrices $\Gamma_i$ can be put in a block antidiagonal form,
$\Gamma_i = \left(\begin{array}{cc}
  0 & \sigma_i \\
  {\tilde\sigma}_i & 0
\end{array}\right),$
for $i=1,...,N$. The correspondence is such that the
space-dimensionality coincides with $N$, the number of extended
supersymmetries, while the dimensionality of the $\Gamma_i$
matrices is $2d$, the total number of bosons and fermions.
\par
The reality condition imposed on the $\Gamma_i$ is in consequence
of the reality property of the bosonic and fermionic fields, while
the Weyl property can be understood by loosely stating that the
Clifford's $\Gamma_i$ matrices are now promoted to be
supermatrices which act on and exchange the fermionic and the
bosonic subspaces.\par The classification of all finite real
representations of Clifford algebras, satisfying the further Weyl
property, has been given in the literature in several places (see
\cite{PaTo} for references).\par While the construction in
\cite{PaTo} has been performed independently of and without any
mention to division algebras, just for the special case of
$C(0,8)$ and $C(8,0)$, something more can be said. They are both
Weyl-type representations which are associated with the
$16$-dimensional ($8$ bosons and $8$ fermions) irreducible
fundamental multiplet of the $N=8$ supersymmetry. It is therefore
possible to combine the results of \cite{PaTo} with the relation,
presented in \cite{Oku}, on the connections with such real
Clifford representations and the octonions. The final result
consists in the established connection between a matrix
representation of the global $N=8$ supersymmetry, and the $N=8$
octonionic construction which has been discussed in the previous
section.

\section{SUPERAFFINE ALGEBRAS OVER DIVISION ALGEBRAS}

In the previous section we have seen that there are two related
ways of implementing an $N=8$ global supersymmetry. In terms of a
matrix representation or by making use of the octonions. In this
section I discuss under which conditions the octonionic-realized
$N=8$ global supersymmetry can be recovered from charges of an
associated supercurrent algebra. More generally, I discuss how it
is possible to ``superaffinize" a division algebra, following
\cite{CRT} .\par There exists a well-defined procedure to
(super)affinize any given simple Lie or super-Lie algebra ${\cal
G}$. The details are given e.g. in \cite{Top}, where some physical
applications are also discussed. No matter which is the starting
(super)Lie algebra ${\cal G}$, the associated superaffine algebra
is an ordinary superLie algebra which in particular satisfies the
superJacobi identity. A drastically new feature arises when the
starting algebra is the division algebra of octonions. Due to the
alternative property, the commutator algebra of octonions
\begin{eqnarray}
\relax [\tau_a,\tau_b]&=&f_{abc}\tau_c, \end{eqnarray} (with
$f_{abc}=2C_{abc}$, the structure constants introduced in section
$2$) no longer satisfies the Jacobi identity, but the weaker
condition
\begin{eqnarray}
\relax J(x,y,[x,z]) &=& [J(x,y,z),x], \label{malcev}
\end{eqnarray}
which, together with $\relax [x,x]=0$, gives it the structure of a
Malcev algebra (the property (\ref{malcev}) holds for any three
given elements $x,y,z$, where $J(x,y,z)$ is the jacobian).
\par
Let us consider the algebra of commutators satisfied by the seven
imaginary octonions plus the identity ${\bf 1}$ which, for our
scopes, can also be expressed as $\tau_0$. It admits a
quaternionic subalgebra which coincides with the $sl(2)\oplus
u(1)$ Lie algebra. The superaffinization requires at first the
construction of a superloop algebra. This implies to associate to
any bosonic generator $\tau_i$ a corresponding fermionic $N=1$
superfield $\Psi_i(X)=\psi_i(x) +\theta j_i(x)$. Let us denote
with $f_{ijk}$ the original structure constants $f_{ijk}$, the
superloop algebra can be introduced through
\begin{eqnarray} \{
\Psi_i(X_1),\Psi_j(X_2)\}&=&
f_{ijk}\Psi_k(X_2)\delta(X_1,X_2),\nonumber\\ && \label{superloop}
\end{eqnarray}
(where $\delta(X_1,X_2)\equiv \delta(x_1-x_2)(\theta_1 -\theta_2)$
is the supersymmetric delta-function). Next, in order to affinize
the superloop algebra, a central extension $k$ must be introduced
in a manifestly supersymmetric form. In the case of a Lie algebra
(such as the $sl(2)\oplus u(1)$ subalgebra) a central extension
appearing in the r.h.s. of (\ref{superloop}) takes the form
\begin{eqnarray} k\cdot tr(\tau_i\tau_j)D_{X_2}\delta(X_1,X_2),&&
\end{eqnarray} where $D$ is the $N=1$ fermionic derivative
(\ref{derivative}), while the trace is computed on some matrix
representation for $\tau_i,\tau_j$. In the case of octonions, due
to the obvious absence of matrix representations, the above
formula is meaningless and must be replaced by some
generalization. It turns out that the correct generalization is
given by
\begin{eqnarray}
k\cdot \Pi(\tau_i\cdot\tau_j)D_{X_2}\delta(X_1,X_2), &&
\label{centralext}
\end{eqnarray}
where $\Pi(\tau_i\cdot \tau_j)$ corresponds to the projection over
the identity ${\bf 1}$. On the quaternionic subalgebra it
reproduces the correct formula for the central extension, as it
should be.\par The affinization of (the commutator of) the
division algebra of the octonions has been considered in a series
of papers by Osipov, two of them \cite{Osi1} devoted to the
bosonic affine case and its related Sugawara construction, while
the third one \cite{Osi2} to construct the superaffine algebra.
While in the bosonic subalgebra case (setting all fermionic fields
equal to zero) our results coincide with the Osipov's results, the
central extension presented in \cite{Osi2} is not manifestly
supersymmetric, as an easy inspection shows. The central extension
(\ref{centralext}) given here, on the contrary, is manifestly
supersymmetric. We therefore regard the (\ref{superloop}) loop
algebra, centrally extended through (\ref{centralext}) as the
correct superaffinization of the commutator algebra of the
octonions. A first property can be immediately stated, namely that
the superaffine octonionic algebra is a superMalcev algebra,
satisfying a graded version of the Malcev property, which is
expressed in terms of the superjacobian.
\par
Despite the fact that the superaffine algebra here introduced is
expressed in terms of manifestly $N=1$ superfields only, its
association with a division algebra suggests that it possesses
extra supersymmetries. This is indeed the case. It is in fact a
global $N=8$ supersymmetric algebra. The sketch of the proof,
following \cite{CRT}, will be given in the next section, where the
supersymmetric Sugawara construction will be discussed. Let us
conclude this section by mentioning that the restriction to the
quaternionic case, once expressed in component fields, corresponds
to the $N=4$ superaffine algebra introduced in \cite{IKT}, in
terms of a manifest $N=2$ superfields formalism. The division
algebra properties were assumed, but no explicit use of them was
made in that construction.

\section{THE SUGAWARA CONSTRUCTION FOR THE NON-ASSOCIATIVE $N=8$
SCA}

The superaffine algebra introduced in the previous section, which
can be denoted as the ${\hat {\bf O}}$-superalgebra, is $N=8$, as
already recalled. An explicit way to prove it consists in
constructing a Sugawara realization which produces an $N=8$
extension of the Virasoro algebra. This has been done in
\cite{CRT} by using an improved version (to deal with octonions)
of the Thieleman's Mathematica package for OPE's computations. The
$N=8$ extension of Virasoro coincides with the $N=8$ (so-called
non-associative) Superconformal algebra introduced in
\cite{ESTvPS}, which is some sort of minimal $N=8$ extension of
Virasoro, being generated, besides the Virasoro field $T$, by $8$
spin-$\frac{3}{2}$ fermionic fields $Q_i$ and $7$ spin-$1$ bosonic
currents $J_a$. It is constructed with the structure constants of
the octonions and for that reason the Jacobi identities are not
satisfied. \par The Sugawara construction requires the notion of
the enveloping algebra. It is worth stressing the fact that in the
case of the superMalcev ${\hat{\bf O}}$ algebra the bosonic and
fermionic component fields are assumed to be real-valued (and not
octonionic-valued). The ${\hat{\bf O}}$ algebra must therefore be
regarded as an abstract non-Lie algebra of Poisson-Malcev
brackets. When considering the enveloping algebra the
Poisson-Malcev brackets are assumed to satisfy the graded version
of the Leibniz rule. One important remark is the following, the
Malcev property is not closed under composition law, i.e., even if
the triples $x_1,y,z$ and $x_2,y,z$ both satisfy (\ref{malcev}),
it is not guaranteed that the triple $x=x_1 x_2, y,z$ is Malcev.
Indeed, the $N=8$ non-associative SCA is not super-Malcev since,
e.g., the (\ref{malcev}) property is violated for $J_1,J_2,Q_4$,
while its bosonic subalgebra is of Malcev type.
\par
For what concerns the Sugawara construction of ${\hat{\bf O}}$,
two distinct limiting subcases have to be distinguished. The
purely bosonic limit, obtained by disregarding all fermionic
fields, which reproduces the results obtained in \cite{Osi1}, and
the quaternionic case, obtained by disregarding the fields
associated to the values $i=4,5,6,7$. It corresponds to the
Sugawara construction of \cite{IKT}, which reproduces the minimal
$N=4$ Superconformal algebra. Such a Sugawara realizes an
interpolation between the super-NLS and the super-mKdV
hierarchies.\par Since the details of the construction and the
full list of formulas can be found in \cite{CRT}, here I limit
myself to stress the salient points. Unlike the bosonic and the
$N=4$ subcases, the construction of the $N=8$ Sugawara requires a
classical renormalization procedure and is recovered only in the
limit $k\rightarrow \infty$, where $k$ is the central charge (of
the superaffine algebra). The reason is due to the fact that the
algebra of the Sugawara-constructed fields does not close
automatically on the non-associative $N=8$ SCA, since
extra-contributions, proportional to the fermionic fields
$\psi_i(x)$, for $i=4,5,6,7$, appear. These extra-contributions
however disappear in the limit $k\rightarrow\infty$. A
simultaneous rescaling (proportional to $k$) of the Poisson-Malcev
brackets normalization is needed in order to produce a finite
value of the central extension $c$ of the $N=8$ SCA. Due to the
presence of this ``classical renormalization", no relation is
found between the affine central charge $k$ and the conformal
central charge $c$. The latter, with a finite rescaling of the
fields and of the brackets, can be normalized at will. Taking the
$k\rightarrow\infty$ limit is safe since the terms which disappear
in the r.h.s. are always of higher order in $\epsilon=\frac{1}{k}$
w.r.t. the surviving terms which close the $N=8$ SCA.\par To
illustrate the above points I present here the form of the
Sugawara just for the Virasoro field $T(x)$ and the
spin-$\frac{3}{2}$ field $Q(x)$ associated to the identity ${\bf
1}$. We have (the convention over repeated indices is understood)
\begin{eqnarray}
T &=& \frac{1}{k^2}(j_ij_i +{\psi_i}'\psi_i) +\frac{1}{k}{j_0}'
-\nonumber\\ &&\frac{2}{3 k^3}C_{abc}\psi_a\psi_bj_c -
\frac{2}{k^4}C_{abcd}\psi_a\psi_b\psi_c\psi_d,\nonumber\\ Q&=&
\frac{1}{k^2}\psi_ij_i+\frac{1}{k}{\psi_0}'-\frac{2}{3
k^3}C_{abc}\psi_a\psi_b\psi_c,
\end{eqnarray}
where the dual tensor (in the seven-dimensional imaginary
octonionic space) $C_{abcd}= \epsilon_{abcdefg}C_{efg}$ is totally
antisymmetric.
\par
While the Virasoro algebra for $T(x)$ is satisfied exactly, the
spin-$\frac{3}{2}$ primary condition for the $Q_0(x)\equiv Q(x)$
field is guaranteed only in the $k\rightarrow\infty$ limit, indeed
\begin{eqnarray}
\{T(x),T(y)\}_R &=& -\frac{1}{2}\delta''' +2 T(y)\delta' +
T(y)'\delta,\nonumber\\ \{Tx),Q(y)\}_R &=& \frac{3}{2}Q(y)\delta'
+ Q(y)'\delta+ X_1,\nonumber\\ \{Q(x),Q(y)\}_R &=&
-\frac{1}{6}\delta'''+\frac{1}{2}T(y)\delta +X_2,
\end{eqnarray}
(in the above formulas $\delta\equiv\delta(x-y)$ and the
derivative is taken in $y$). The extra-terms $X_1,X_2$ are given
by
\begin{eqnarray}
X_1&=&- \frac{3}{2 k^4}C_{abcd}\psi_a\psi_b\psi_cj_d \delta,
\nonumber\\ X_2 &=& -\frac{5}{2 k^4}
C_{abcd}\psi_a\psi_b\psi_c\psi_d\delta. \end{eqnarray} Notice that
the extra-terms in the r.h.s. are not present when the Sugawara is
restricted to the superaffinization of the quaternionic subalgebra
or in the purely bosonic case.\par Relations like those here shown
hold for the whole set of fields which satisfy the $N=8$
non-associative superconformal algebra.\par By taking the Poisson
brackets between the supersymmetric charges $Q_i=\int dx Q_i(x)$
and the original affine fields in ${\hat{\bf O}}$, we can recover
in the $k\rightarrow\infty $ case the $N=8$ ``octonionic"
transformations (\ref{transf1}) and (\ref{transf2}), assumed to
act on the octonionic-valued superfield $\Psi_i(X)\tau_i$. We
have, e.g.,
\begin{eqnarray}
\delta_{Q_1} j_1 &=& \frac{1}{2}{\psi_0}'
+\frac{2}{k^2}(C_{abc}\psi_a\psi_b\psi_c), \nonumber\\
\delta_{Q_1} \psi_1 &=& -\frac{1}{2}j_0,\nonumber\\ \delta_{Q_1}
j_2 &=& \frac{1}{2}{\psi_3}' +O(\frac{1}{k}),\nonumber\\
\delta_{Q_1}\psi_2 &=& -\frac{1}{2}j_3 +\frac{1}{k}(\psi_1\psi_2
-\psi_0\psi_3). \label{sutr}
\end{eqnarray}
(in the third equation the term $O(\frac{1}{k})$ is rather
complicated and is not necessary to report it here). Analogous
results are found for the whole set of supersymmetric
transformations $\delta_{Q_i}$ acting on the fields $\psi_k(x),
j_k(x)$. Due to the ``democratic" nature of the octonions (they
are all on equal footing), the most general supersymmetry
transformation can be directly read from (\ref{sutr}).\par The
global $N=8$ octonionic transformation is interpreted, as
promised, in terms of a supercurrent algebra.

\section{AN APPLICATION: DIVISION ALGEBRAS AND THE $N$-EXTENDED
SUPERKdV}

From a practical point of view one of the most interesting
features of the non-associative $N=8$ SCA is the presence of the
central charge. The fact that this superalgebra does not satisfy
the (super)Jacobi property allows to overcome the no-go theorem
which states that central charges are not allowed for
superconformal algebras with $N>4$. This is the case for standard
superalgebras which satisfy the (super)Jacobi property. On the
other hand the presence of a central charge is required if the
algebra under consideration has to be interpreted as a
Poisson-bracket algebra for a related KdV-type hamiltonian
equation (the Virasoro central charge $\delta'''$ produces the
$T'''$ term in the usual ${\dot T}=T'''+ T T'$ KdV equation). It
is therefore interesting to explore the possibility that the
superaffine division algebra ${\hat{\bf O}}$ (and its subalgebras)
could be related to supersymmetric extensions of KdV, generating a
hamiltonian dynamics. This possibility has been systematically
investigated in \cite{CRT2}, with a heavy use of computer
algebraic manipulations. Leaving aside for the moment all
questions related with the integrability of the extended system,
in \cite{CRT2} the most general $N=4$ and $N=8$ hamiltonian
supersymmetric extensions of KdV, based on the $N=4$ minimal SCA
and $N=8$ non-associative SCA respectively, have been constructed.
The complete solution is the following. The most general
$N=4$-supersymmetric hamiltonian for the $N=4$ KdV depends on $5$
parameters (plus an overall normalization constant). However, if
the hamiltonian is further assumed to be invariant under the
involutions of the $N=4$ minimal SCA, three of the parameters have
to be set equal to $0$. I recall here that the involutions of the
$N=4$ minimal SCA are induced by the involutions of the
quaternionic algebra. Three such involutions exist (any two of
them can be assumed as generators), the $a$-th one (for $a=1,2,3$)
is given by leaving $\tau_a$ (together with the identity)
invariant and flipping the sign of the two remaining $\tau$'s.\par
What is left is the most general hamiltonian, invariant under
$N=4$ and the involutions of the algebra. It is given by
\begin{eqnarray}
H_2 &=& T^2 +Q_i'Q_i -J_a''J_a +x_a T{J_a}^2 +\nonumber\\ && 2x_a
Q_0Q_aJ_a -\epsilon_{abc}x_aJ_a Q_bQ_c +\nonumber\\ && 2
x_1J_1J_2'J_3-2x_2J_1'J_2J_3
\end{eqnarray}
(the convention over repeated indices is understood; here
$i=0,1,2,3$ and $a=1,2,3$).
\par
The $N=4$ global supersymmetry requires the three parameters $x_a$
to satisfy the condition
\begin{eqnarray}
x_1+x_2+x_3&=&0,
\end{eqnarray}
so that only two of them are independent. Since any two of them,
at will, can be plotted in a real $x-y$ plane, it can be proven
that the fundamental domain of the moduli space of inequivalent
$N=4$ KdV equations can be chosen to be the region of the plane
comprised between the real axis $y=0$ and the $y=x$ line
(boundaries included). There are five other regions of the plane
(all such regions are related by an $S_3$-group transformation)
which could be equally well chosen as fundamental domain.\par In
the region of our choice the $y=x$ line corresponds to an extra
global $U(1)$-invariance, while the origin, for $x_1=x_2=x_3=0$,
is the most symmetric point (it corresponds to a global $SU(2)$
invariance).
\par
The involutions associated to each given imaginary quaternion
allows to consistently reduce the $N=4$ KdV equation to an $N=2$
KdV, by setting simultaneously equal to $0$ all the fields
associated with the $\tau$'s which flip the sign, e.g. the fields
$J_2=J_3=Q_2=Q_3=0$ for the first involution (and similarly for
the other couples of values $1,3$ and $1,2$). After such a
reduction we recover the $N=2$ KdV equation depending on the free
parameter $x_1$ (or, respectively, $x_2$ and $x_3$).
\par The integrability is known for $N=2$ KdV to be ensured for three
specific values $a=-2,1,4$, discovered by Mathieu \cite{Mat}, of
the free parameter $a$. We are therefore in the position to
determine for which points of the fundamental domain the $N=4$ KdV
is mapped, after any reduction, to one of the three Mathieu's
integrable $N=2$ KdVs. It turns out that in the fundamental domain
only two such points exist. Both of them lie on the $y=x$ line.
One of them produces, after inequivalent $N=2$ reductions, the
$a=-2$ and the $a=4$ $N=2$ KdV hierarchies. The second point,
which produces the $a=1$ and the $a=-2$ $N=2$ KdV equations,
however, does not admit at the next order an $N=4$ hamiltonian
which is in involution w.r.t. $H_2$. In this respect the results
in \cite{CRT2}, which can be regarded as more general since are
based on an exhaustive component-fields analysis, are in agreement
with those of \cite{DeIv}, based on an extended superfield
formalism.\par The next step considered in \cite{CRT2} was to
perform a similar analysis for the $N=8$ case based on the $N=8$
non-associative SCA. At first the most general hamiltonian with
the right dimension has been written down. Later, some constraints
on it have been imposed. The first set of constraints requires the
invariance under all the $7$ involutions of the algebra. In the
case of octonions the total number of involutions is $7$ (with $3$
generators) each one being associated to one of the seven
combinations appearing in (\ref{combination})). In the case, e.g.,
of $123$ the corresponding $\tau_a$'s are left invariant, while
the remaining four $\tau$'s, living in the complement, have the
sign flipped.
\par The second set of constraints requires the invariance under
the whole set of $N=8$ global supersymmetries. Under this
condition there exists only one hamiltonian, up to the
normalization factor, which is $N=8$ invariant. It does not
contain any free parameter and is quadratic in the fields. It is
explicitly given by
\begin{eqnarray}
H_2 &=& T^2 + Q_i'Q_i - J_a''J_a \label{quadham}
\end{eqnarray}
(here $i=0,1,...,7$ and $a=1,2,...,7$).\par The hamiltonian
corresponds to the origin of coordinates (confront the previous
case) which is also, just like the $N=4$ case, the point of
maximal symmetry. This means that the hamiltonian is invariant
under the whole set of seven global charges $\int dx J_a(x)$,
obtained by integrating the currents $J_a$'s.
\par Despite its apparent simplicity, it gives
an $N=8$ extension of KdV which does not reduce (for any $N=2$
reduction) to the three Mathieu's values for integrability.
Nevertheless the main result, highly non-trivial, consists in the
proof of the existence of an $N=8$ extension of the KdV equation
whose dynamics is hamiltonian (for a generalized non-Lie
Poisson-brackets structure).\par As a matter of fact, requiring
the invariance under $N=5$ supersymmetries already uniquely
specifies (\ref{quadham}), since the remaining supersymmetries are
generated by the previous ones. $N=4$ is the maximal supersymmetry
which allows the presence of a global parameter. Two inequivalent
ways of determining a free parameter-dependent $N=4$-invariant
second hamiltonian $H_2$, constructed with the $N=8$ SCA fields,
are allowed. The first one requires $N=4$ invariance w.r.t. the
$N=4$ global supercharges constructed from the $N=4$ minimal SCA
subalgebra. The second one requires the invariance w.r.t. the
remaining $4$ supersymmetric generators. The explicit formulas for
such $H_2$'s and the derived hamiltonian equations are presented
in \cite{CRT2}.

\section{CONCLUSIONS}

In this paper I have presented some new results concerning an
explicit connection of extended supersymmetries and division
algebras. Besides the relation between matrix and octonionic
representations of the global $N=8$ supersymmetry, the
superaffinization of a Malcev algebra based on octonions has been
introduced, as well as its $N=8$ Sugawara construction which links
it to the non-associative $N=8$ SCA of reference
\cite{ESTvPS}.\par As an application, the role of such algebras in
furnishing a generalized Poisson bracket structure for the
hamiltonian dynamics of supersymmetric extensions of the KdV
equation has been examined. It has been proven, following
\cite{CRT2}, the existence of a unique $N=8$ KdV equation based on
the $N=8$ non-associative SCA. Many more applications are likely
to be studied with the tools here discussed. A list of possible
topics can be furnished. In the light of the geometric approach,
see \cite{BSV}, it is likely that the dynamics of a classical
superstring moving in a flat target could be recovered from a
supersymmetric version of the Liouville equation. If the target is
$3$-dimensional, this is indeed the case. For $4$ dimensional, $6$
dimensional and $10$-dimensional targets the corresponding
super-Liouville theory is expected to take values in the complex,
quaternionic and octonionic division algebras respectively. The
algebraic structures here discussed seem the most natural
framework to investigate such questions. On the other hand, either
WZNW-type of models defined on the ``almost group manifold" $S^7$
(the almost being referred to the fact that it can be recovered
from imaginary octonions, together with their product which,
however, is non-associative), or the dynamics of generalized tops
moving on the $S^7$ sphere, are likely to be described by Malcev
algebras. A converse approach, which retains Jacobi identities at
the price of making field-dependent the algebraic structure
constants (soft algebras) has been suggested in \cite{CePr}.\par
Another natural setting concerns the application of such formalism
to the twistors in investigating the quantization of the
Green-Schwarz string \cite{Ber}.

\par

{\bf Acknowledgments}

I am grateful to the organizers of the Kharkov International
Conference ``Supersymmetry and Quantum Field Theory" for their
kind invitation. It is a pleasure to thank my collaborators A.
Pashnev from Dubna and H.L. Carrion and M. Rojas from CBPF.

\end{document}